\documentclass[twocolumn,preprintnumbers,amsmath,amssymb,superscriptaddress]{revtex4}

\usepackage{graphicx}
\usepackage{dcolumn}
\usepackage{bm}
\usepackage{color}
\usepackage{ulem}
 \usepackage{subfigure}

\begin{document}

\title{Optical Random Riemann Waves in Integrable Turbulence}

\author{St\'ephane Randoux}
 \email{stephane.randoux@univ-lille1.fr}
\affiliation{Univ. Lille, CNRS, UMR 8523 - PhLAM -
 Physique des Lasers Atomes et Mol\'ecules, F-59000 Lille, France}
\author{Fran\c{c}ois Gustave}
\affiliation{Univ. Lille, CNRS, UMR 8523 - PhLAM -
 Physique des Lasers Atomes et Mol\'ecules, F-59000 Lille, France}
\author{Pierre Suret}
\affiliation{Univ. Lille, CNRS, UMR 8523 - PhLAM -
 Physique des Lasers Atomes et Mol\'ecules, F-59000 Lille, France}
\author{Gennady El}
\affiliation{Centre for Nonlinear Mathematics and Applications, Department of Mathematical Sciences, Loughborough
 University, Loughborough LE11 3TU, United Kingdom}

\date{\today}

\begin{abstract}

We examine integrable turbulence (IT) in the framework of the defocusing cubic one-dimensional nonlinear
Schr\"{o}dinger equation. This is done theoretically and experimentally, by realizing an optical fiber
experiment in which the defocusing Kerr nonlinearity strongly dominates linear dispersive effects.
Using a dispersive-hydrodynamic approach, we show that the development of IT can be divided
into two distinct stages, the initial, pre-breaking stage being described by a system of interacting random
Riemann waves. We explain the low-tailed statistics of the wave intensity in IT and show that the
Riemann invariants of the asymptotic nonlinear geometric optics system represent the observable
quantities that provide new insight into statistical features of the initial stage of the IT development
by exhibiting stationary probability density functions.
\end{abstract}


\maketitle

Propagation of nonlinear random waves has recently received much
attention  in many areas of modern physics such as nonlinear
statistical optics \cite{Picozzi:14,Laurie:12,Turitsyna:13,Pierangeli:16},
hydrodynamics \cite{Herbert:10}, mechanics \cite{Miquel:13},  and
cold-atom physics \cite{Navon:16}.  In all these areas a broad class of wave phenomena
is modelled by integrable  nonlinear partial differential equations (PDEs).  Although the fundamental role of
integrable PDEs has been established since the pioneering work of
Fermi, Pasta and Ulam in the $1950$s \cite{FPU:55}  the significance
of random input problems for such systems was realized only
recently,  leading to the concept of {\it integrable turbulence (IT)}
\cite{Zakharov:09,Agafontsev:15,Randoux:14,Walczak:15,Suret:16,Nahri:16,Randoux:16b,Zakharov:16,Agafontsev:16}.
In this context, the one-dimensional nonlinear Schr\"{o}dinger
equation (1D-NLSE) plays a prominent role because it describes at leading 
order wave phenomena relevant to many fields of nonlinear physics.  

It is now well established from experiments and numerical simulations
that  heavy-tailed (resp. low-tailed) deviations from gaussian
statistics occur in integrable wave systems  ruled by the focusing
(resp. defocusing)  1D-NLSE
\cite{Walczak:15,Suret:16,Randoux:14,Randoux:16b}.  The heavy-tailed
deviations from gaussian statistics have their origin in the random
formation  of bright coherent structures having properties of
localization in space and time  similar to rogue waves
\cite{Walczak:15,Suret:16,Onorato:13}. On the other hand, the
low-tailed deviations are due to  random generation of dispersive
shock waves (DSWs) and dark solitons \cite{Randoux:14,Randoux:16b}.
One of the key features of IT is the establishment, at long evolution
time, of a state in which the statistical properties of the wave
system remain stationary.  Due to integrable nature of the system, the
long-time statistics depends on the statistics of the input random
process (cf. \cite{Agafontsev:15,Randoux:14,
  Walczak:15,Randoux:16b}). So far, there has been no satisfactory
theoretical framework developed for the description of statistical
features of IT due to  high complexity of nonlinear wave interactions
occurring over the course of its development.

In this Letter, we examine IT in optical systems described by the
defocusing 1D NLSE from the perspective of dispersive hydrodynamics
\cite{Biondini:16}, a semi-classical theory of nonlinear dispersive
waves exhibiting two distinct spatio-temporal scales: the long scale
specified by initial conditions and the short scale by the internal
coherence length (i.e. the typical size of the  coherent structures).
This scale separation enables one to split the development of IT into
distinct stages characterized by qualitatively different
dynamical and statistical features.

At the initial, pre-breaking stage of dispersive-hydrodynamic IT
nonlinear effects dominate linear dispersion and the wave fronts of
the random initial field experience gradual  steepening leading to the
formation of gradient catastrophes that are  regularized  through the
generation of  DSWs \cite{el:2016}.  As  numerical simulations
reported in ref. \cite{Randoux:16b} show, the pre-breaking stage of IT
is characterized by significant  deviations from the gaussian
statistics  exhibiting the  low-tailed probability density function
(PDF) for the wave's intensity, see also Fig. 1(b).  In the post-breaking regime, the
evolution of the statistics of the random wave field is determined by
interactions among DSWs leading to further deviations from gaussianity
(see ref. \cite{Randoux:16b} and Fig. 1(c) showing the PDFs 
before the occurrence of wave breaking ($\xi=0.156$) and at long 
evolution distance ($\xi=1.56$), in the statistical stationary state).

In this work, we provide a quantitative explanation of the occurrence
of non-gaussian statistics at the pre-breaking stage of the  IT
development  by analysing solutions of  the defocusing 1D-NLSE in the
zero dispersion (nonlinear geometric optics) limit, where the dynamics can be
interpreted in terms of interacting {\it Riemann waves}.  Moreover, we
show that {\it Riemann invariants} diagonalising the geometric optics
system  represent also the relevant  statistical variables  in IT,
exhibiting {\it stationary PDFs}, in sharp contrast  with evolving
statistical distributions of the instantaneous power.

We consider the defocusing integrable 1D-NLSE
in dimensionless form:
\begin{equation}\label{nlse}
i \epsilon \frac{\partial \psi}{\partial \xi} + \frac{\epsilon^2}{2}
\frac{\partial^2 \psi}{\partial \tau} -|\psi|^2 \psi=0.
\end{equation}
In the optical fiber experiment realized in our work, 
$\psi(\xi,\tau)$ is the slowly-varying envelope of
the electric field $A$ that  is normalized to the square root of the
mean optical power $\bar{\rho_0}$ of the partially  coherent field
propagating inside the fiber ($\psi=A/\sqrt{\bar{\rho_0}}$). It is usual  in
nonlinear fiber optics to introduce a nonlinear length
$L_{NL}=1/(\gamma \bar{\rho_0})$ and a linear  dispersion length
$L_D=2/(\beta_2 [\Delta \nu_0]^2)$. $\gamma$ and $\beta_2$ are the Kerr
and the second-order dispersion
coefficients of the optical fiber, respectively  ($\beta_2=+20$ps$^2$km$^{-1}$,
$\gamma=+6$W$^{-1}$km$^{-1}$,  normal dispersion regime).  $\Delta
\nu_0$ represents the width of the spectrum of power fluctuations of
the wave injected inside the optical
fiber. With our notations, the propagation distance $z$ along the
fiber is normalized as $\xi=z/\sqrt{L_{NL}L_D}$, the physical time $t$
is normalized as $\tau=t/T_0$ with $T_0=1/\Delta \nu_0$ and
$\epsilon=\sqrt{L_{NL}/L_D}$ is the dispersion parameter which in our experiment is about $0.014$.

Considering $\epsilon$ to be a small parameter we introduce the  semi-classical Madelung transformation
$\psi(\xi,\tau)=\sqrt{\rho(\xi,\tau)}e ^{i
  \frac{\phi(\xi,\tau)}{\epsilon}}$   to obtain to leading order in
$\epsilon$  the nonlinear geometric optics equations for the the instantaneous
power  $\rho(\xi,\tau)$  and  the  instantaneous frequency  $u(\tau,
\xi)= \phi_\tau$ of the optical wave
\cite{forest_exact_2009,Wabnitz:13b,Fatome:14,Kodama1995,Moro:14}
\begin{equation}\label{shallow}
    \rho_{\xi}+(\rho u)_{\tau}=0, \qquad u_{\xi}+uu_{\tau}+\rho_{\tau}
    =0 \, .
\end{equation}
Eqs. (\ref{shallow}) are identical to the shallow-water
equations for an incompressible fluid  with $\rho>0$ and $u$ interpreted as the
fluid height and  the depth-averaged horizontal fluid
velocity respectively and with the roles of space $\xi$ and
time $\tau$ interchanged. System (\ref{shallow}) was rigorously proved in \cite{jin:1999} to describe the pre-breaking
NLS dynamics in the semi-classical  ($\epsilon \to 0$) limit. 

Upon introducing Riemann invariants $r_{1,2}(\xi,\tau)=u \pm 2 \sqrt{\rho}$
as new variables, 
the system (\ref{shallow}) becomes \cite{el:2016}
\begin{equation}\label{riemann_sw}
\frac{\partial r_{1,2}}{\partial \xi} + V_{1,2} \frac{\partial
  r_{1,2}}{\partial \tau}=0, \quad V_{1,2} = \frac{3}{4} r_{1,2} + \frac{1}{4} r_{2,1}.
  \end{equation}
For non-constant $r_{1,2}$ system (\ref{riemann_sw}) describes 
the propagation of two interacting Riemann waves (RWs) \cite{whitham}. 

Fig. 1 shows a typical result of the numerical integration
of Eq. \ref{nlse} by taking  
a random field $\psi(\xi=0,\tau)$ having  gaussian statistics
as initial condition.  
This random initial field is composed of a 
sum of independent Fourier modes with random phases, see 
Supplemental Material and ref. \cite{Randoux:16b,Nazarenko}.
Fig. 1 reveals the contrasting behaviours of dynamics and statistics of $\rho$ and $r_{1,2}$ 
at the pre-breaking stage of IT.
 
As shown in Fig. 1(a), the front edges of $\rho$ experience 
some steepening while the changes in $\rho$ are more pronounced 
at the points where the random field exhibits local maxima. 
At the same time, Fig. 1(b) reveals that Riemann invariants 
$r_{1,2}$ of the wave system behave as counterpropagating 
random waves, as it can be anticipated from  Eq. (\ref{riemann_sw}).

\begin{figure}[h]
\includegraphics[width=0.5\textwidth]{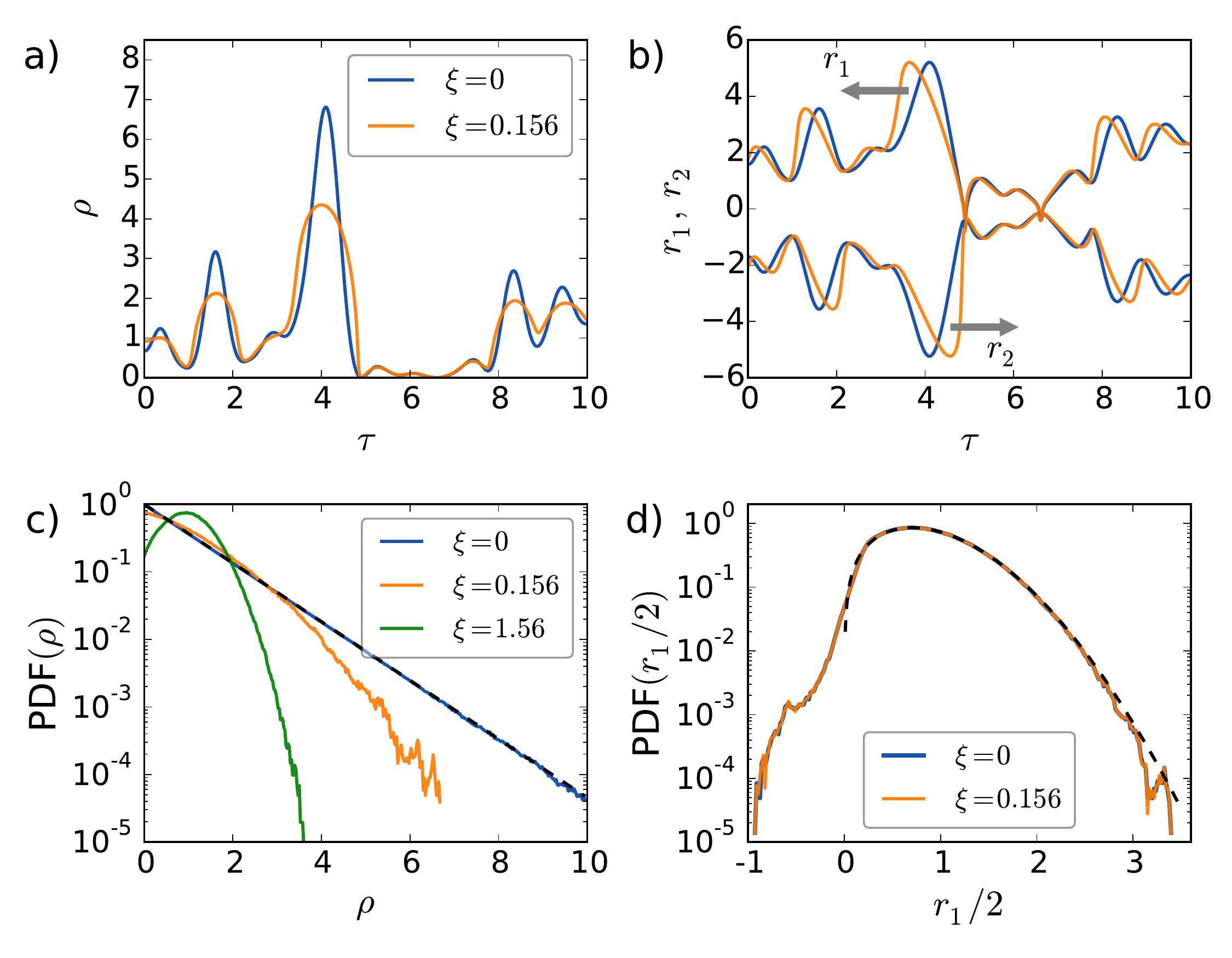}
\caption{Numerical simulation of Eq. (\ref{nlse}) 
 between $\xi=0$ and  $\xi=0.156$ with $\epsilon=0.014$. 
Time evolution of (a) $\rho$, (b) $r_{1,2}$ 
and (c), (d) associated PDFs.  In (c) the black dashed line 
represents the exponential distribution $\mathcal{P}(\rho)=e^{-\rho}$ 
and in (d), it represents the Rayleigh distribution 
($\mathcal{P}(x)=2xe^{-x^2/2}$). The PDF in green line in (c) 
is obtained at $\xi=1.56$ where the wave system has reached 
a statistical stationary state.
}
\end{figure}

The dynamical features evidenced in Fig. 1(a),(b) determine
the statistical properties characterizing IT in the pre-breaking regime. As shown in Fig. 1(c) 
and also previously reported in ref. \cite{Randoux:14,Randoux:16b}, 
the PDF of $\rho$ exhibits low-tailed deviations from the exponential
distribution that arise from changes in $\rho$ seen
in Fig. 1(a).  Contrastingly, the numerical simulations suggest that the PDFs of the Riemann invariants $r_{1,2}$ in IT remain  stationary, despite the noticeable evolution of $r_{1,2}$ themselves (see Fig. 1(b) for the evolution of $r_{1,2}$  and Fig. 1(d) for the PDF of $r_1/2$).
Moreover,  as the initial velocity $u$ is close to zero, the stationary PDFs of $r_{1,2}$ are very close to the Rayleigh distribution shown in Fig. 1(d) by a dashed line.

The contrasting nature of the evolutions of the PDFs of 
$\rho$ and $r_1$ (or $r_2$) evidenced by the numerical simulations 
presented in Fig. 1 represents a striking feature that 
provides a new insight into the initial pre-breaking stage of the development of 
IT. Going beyond numerical simulations, we have used an analytical 
approach to investigate dynamical and statistical features typifying
random Riemann waves. 

First, we show that the dynamical evolution presented in Fig. 1(a) 
can be analyzed from the shallow water equations (\ref{shallow}) with random
initial conditions $\rho(\tau, 0)=\rho_0(\tau)$, $u(\tau, 0)= u_0(\tau)$, 
whose statistics is defined by the input Gaussian process $\psi(\tau, 0)$. 
In our typical experimental and numerical input data, 
we have $u_0(\tau) \ll \sqrt{\rho_0(\tau)}$ so we shall be
assuming $u_0(\tau)=0$ in the analytical development. Looking for the
asymptotic solution of (\ref{shallow}) in the form of ``short-time''
expansions  for  $\rho(\tau, \xi)$ and $u(\tau, \xi)$ we readily
obtain for $\xi \ll 1$:
\begin{equation}\label{analytic_rho_u}
\rho(\tau,\xi) \simeq \rho_0(\tau) + \tfrac14{[\rho_0^2(\tau)]'' \xi^2} ,  \  u(\tau,\xi) \simeq - \rho'_0(\tau) \xi \, .
\end{equation}
 At the points $\tau_m$ of local maxima of $\rho_0(\tau)$
 we have $[\rho_0^2(\tau_m)]'' = 2\rho_0(\tau_m)[\rho_0(\tau_m)]'' <0$ so that the first equation in
 (\ref{analytic_rho_u})  immediately implies the appearance of low tails in 
the PDF of $\rho$ due to the decrease of the maximum amplitude of 
the peaks of fluctuations of $\rho$ with  the evolution
variable $\xi$ (note that in the focusing case  the expansion of $\rho$ has the same form (\ref{analytic_rho_u}) but with the minus sign for the $O(\xi^2)$ term explaining the heavy-tailed statistics observed in \cite{Walczak:15}).

The stationary nature of the PDF of $r_1$ evidenced in Fig. 1(d) 
can be analyzed from Eq. (\ref{riemann_sw})
by noticing that the condition $u \ll \sqrt{\rho}$  must be
satisfied at least over some propagation distance since
$u_0(\tau) \ll \sqrt{\rho_0(\tau)}$. We obtain that in
 the regime of our interest $V_{1,2} \approx \frac12 r_{1,2}$ 
and Eqs. (\ref{riemann_sw}) can then be approximated  by the
system of two {\it decoupled} RWs
\begin{equation}\label{Riem_approx}
\frac{\partial r_i}{\partial \xi} + \frac{r_i}{2}\frac{\partial
  r_i}{\partial \tau}=0, \quad i=1,2.
\end{equation}
Evolution of statistical parameters of random RWs has been studied in
the context of Burgers turbulence  \cite{gurbatov1991nonlinear}. One
of the  straightforward results of the developed
theory is that the PDF $\mathcal{P}(r; \xi)$ of a random RW field
$r(\tau, \xi)$ is invariant with respect to the $\xi$-evolution,
i.e. $\mathcal{P}(r; \xi) = \mathcal{P}(r; 0) $.  The small $u$
approximation (\ref{Riem_approx}) of the  dispersionless dynamics
(\ref{shallow}) then implies that the PDFs of the Riemann invariants
$r_{1,2}$ in the full NLS equation (\ref{nlse}) will remain stationary or almost
stationary during the initial evolution of IT. 

It should be emphasized that numerical simulations shown 
in Fig. 1 are made for 
the regime in which the condition $u \ll \sqrt{\rho}$ of 
our theoretical analysis is not fullfilled. 
Hence the numerical results of Fig. 1 reveal that the 
conservation of the PDF of Riemann invariants holds
at a much longer (but still pre-breaking) evolution time, when the two RWs are coupled
and their evolution is governed by Eq. (\ref{riemann_sw}) instead 
of Eq. (\ref{Riem_approx}). 
This statistical result represents an important extension  of the random
RWs theory \cite{gurbatov1991nonlinear} 
deserving further theoretical analysis.

Now we report an optical fiber experiment in which we realize 
the first observation of random Riemann waves in a turbulent field.
Before presenting our experimental results, let us 
emphasize that experimental observations of RWs that have been 
reported so far involve the setting implying only 
one `isolated' RW \cite{Wetzel:16,Trillo:16}. 
In the nonlinear optics context, this corresponds to imposing a 
very special relation between the wave intensity and the phase
gradient (chirp) \cite{Wabnitz:13}. Such specially designed
optical RWs have been recently realized in optical fiber 
experiments reported in ref. \cite{Wetzel:16} and the wave 
breaking dynamics of one simple RW has been also examined 
in some recent hydrodynamical experiments \cite{Trillo:16}. 

In the context of IT, the intrinsic random 
nature of nonlinear waves prevents the realization of a simple
setting in which the dynamics of the wave system would be 
given by one Riemann invariant while the other one would remain 
constant. This has major implications for the experiment that
must be designed in order to measure not only one hydrodynamical 
variable but both $\rho$ and $u$ in a simultaneous 
way. Moreover the observation of the changes experienced by the 
random RWs can be made only if $\rho$ and $u$ are 
{\it simultaneously} measured at the input and output 
ends of the nonlinear  medium. 

Fig. 2 represents the experimental setup that we have designed 
to perform the measurement of Riemann invariants in the context 
of IT. A partially-coherent light beam at
$1064$ nm is generated by a homemade source that has a  narrow
linewidth together with a gaussian statistics.  The typical time scale
characterizing power fluctuations of this light  source is $T_0 \sim
250$ ps ($\Delta \nu_0=4$ GHz). The optical power of the 
beam is amplified to $\bar{\rho_0} \sim 130$ mW by using an
Ytterbium fiber amplifier.  The partially coherent light beam is
linearly-polarized and it is launched inside a $1.4$~km-long
polarization-maintaining (PM) single-mode  optical fiber. 
In our experiment, the linear and nonlinear  lengths are
$L_D=6250$ km and $L_{NL}=1.3$ km ($\epsilon \sim 0.014$). The 
normalized propagation distance corresponding to the $1.4$~km physical 
distance is $\xi=0.0156$.

\begin{figure}[h]
\includegraphics[width=0.45\textwidth]{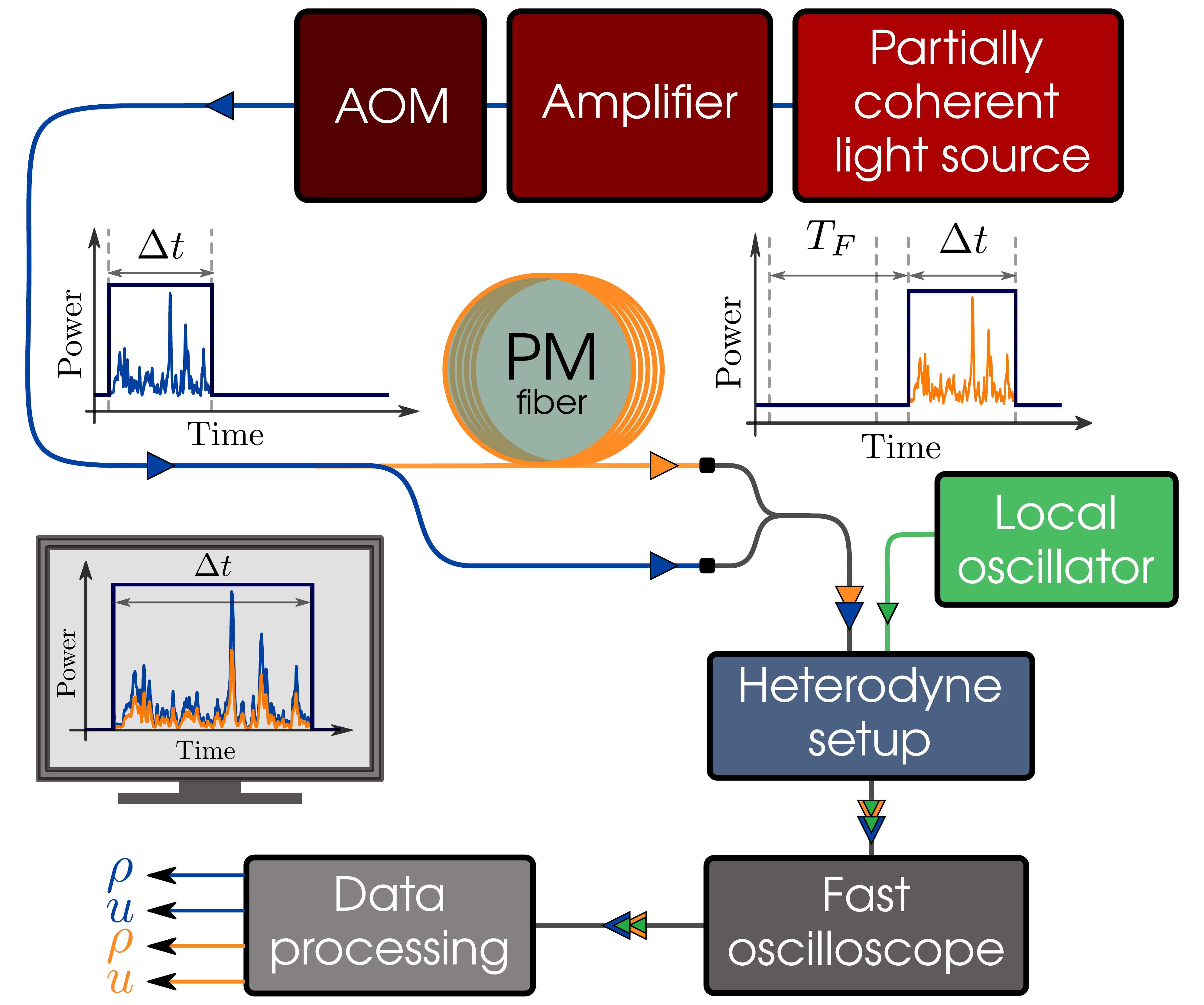}
\caption{Experimental setup. Partially coherent light at $1064$ nm  is
  injected inside a $1.4$-km-long PM fiber in a regime where nonlinear
  effects strongly  dominate linear ones ($L_{NL}=1.3$ km,
  $L_{D}=6250$ km, $\epsilon=0.014$). Real-time observation  of
  $\rho(\tau)$ and of $u(\tau)$ at the input and output ends of the  PM
  fiber is achieved by combining a time-division multiplexing
  technique  and an heterodyne measurement (see text).}
\end{figure}

As shown in Fig. 2, the partially-coherent light wave at the input and
output ends  of the PM fiber is analyzed by using a heterodyne
setup. The light wave is linearly mixed with an external
laser source, also called local oscillator, that delivers stable
single-frequency radiation at $1064$ nm. Two fast photodiodes  having
a bandwidth of $50$ GHz are used in the heterodyne setup
to record the power fluctuations of the incoherent light wave
and the beating signal between the partially-coherent light and 
the local oscillator. The two photodiodes are connected to a fast oscilloscope 
(bandwidth $65$ GHz, sampling rate $160$ GSa/s). Signals detected by
the two photodiodes  have been carefully synchronized with an accuracy
of $\sim 3$ ps  by using a mode-locked laser delivering picosecond
pulses and an adjustable delay line, see Supplemental Material for 
details about the heterodyne measurement of $u$ and the synchronization procedure. 

The experimental setup incorporates
a time-division  multiplexing part that enables the accurate
observation of the nonlinear changes  experienced by $\rho(\tau)$ and
$u(\tau)$ between the input and the output ends of the  PM fiber. 
An acousto-optic modulator (AOM) is used to
periodically slice square windows with a duration $\Delta T=5.7 \mu$s $\gg T_0\sim 250$ps 
in the light wave that is injected inside the PM fiber. 
A $50/50$ fiber coupler is used to combine
light beams at the input and at the output ends of the PM fiber. 
Hence the heterodyne setup periodically analyzes input light fluctuations 
and subsequently, output light fluctuations that are delayed by a time $T_F \sim 7$ $\mu$s 
associated with propagation inside the PM fiber.  Computing the
autocorrelation function of the power fluctuations $P(t)$,  we have
been able to measure $T_F$ with an accuracy of $\pm 3$ ps. 
Data  have been processed in
such a way that light fluctuations at the output  of the fiber are
shifted backward in time by $T_F$, which permits the direct
observation of the nonlinear changes experienced by $\rho(\tau)$ 
and $u(\tau)$ inside the PM fiber. 

\begin{figure}[h]
\includegraphics[width=0.45\textwidth]{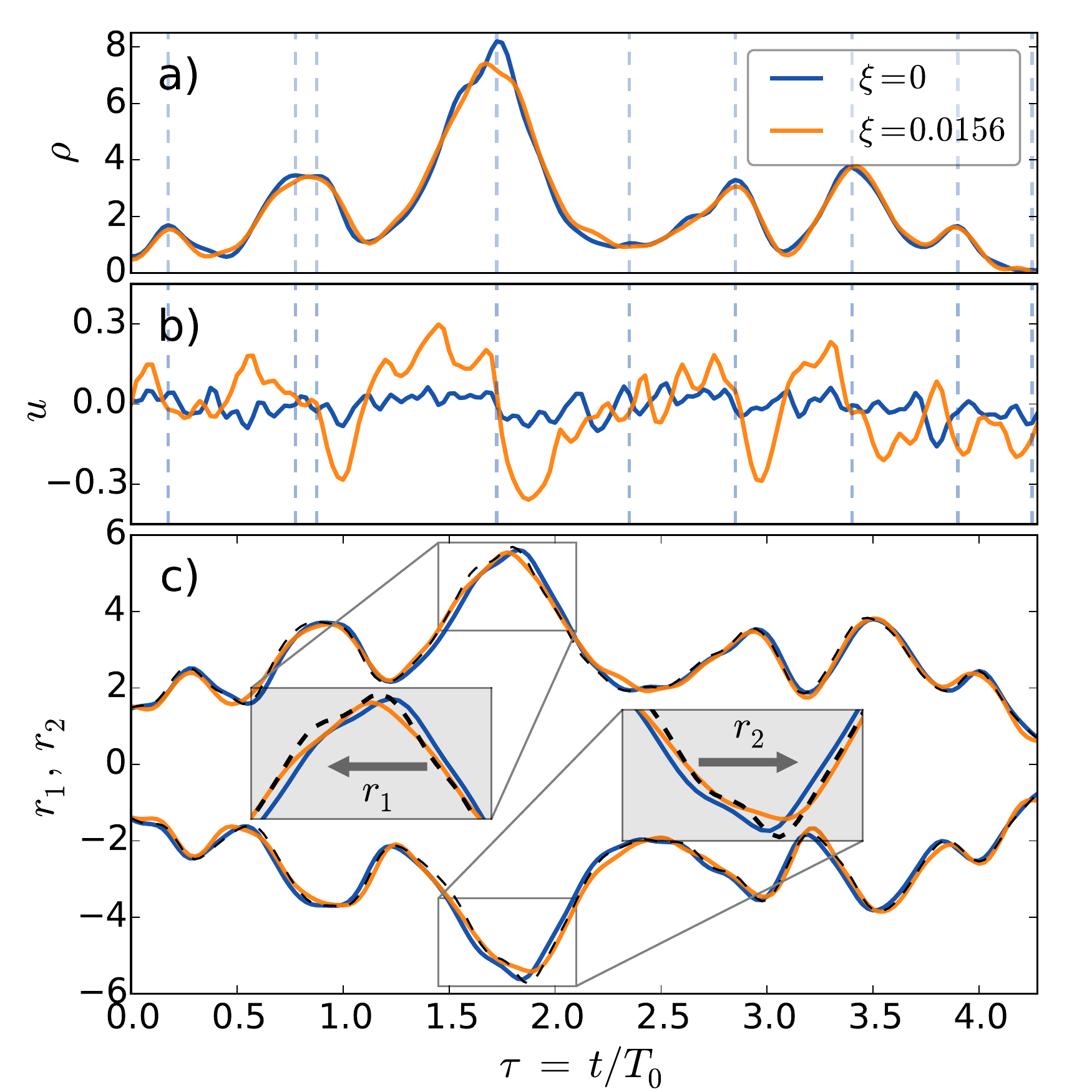}
\caption{Experiments. Time evolution of $\rho$, $u$, and of 
  the Riemann invariants $r_{1,2}=u \pm 2 \sqrt{\rho}$ at the input end ($\xi=0$)
  and at the output end ($\xi=0.0156$) of the fiber
  ($T_0=250$ ps). The black dashed lines represent Riemann invariants
  calculated from numerical integration of Eqs. (\ref{Riem_approx})
  while starting from the experiment initial conditions
 (blue lines).}
\end{figure}

As shown in Fig. 3(a), the experiment reveals dynamical features 
for $\rho(\tau)$ that are similar to 
those evidenced by numerical simulations of Fig. 1(a). 
As shown in Fig. 3(b), the experiment also reveals that the instantaneous 
frequency  $u(\tau)$ does not change
in regions where $\rho(\tau)$ reaches extrema, see vertical dashed
lines in Fig. 3(a)(b) indicating that the positions of maxima of $\rho$
coincide with positions where $u$ stays close to zero. This experimental 
result is in full agreement with the expression obtained for $u$
in Eq. \ref{analytic_rho_u}. 

Fig. 3(c) shows the two Riemann invariants $r_{1,2}=u \pm 2 \sqrt{\rho}$
that are computed from  the data plotted in Fig. 3(a)(b). The
evolution plotted in Fig. 3(c) agrees quite  well with the one
given by Eq. (\ref{Riem_approx}). The Riemann invariants evolve 
as two waves that propagate in opposite directions. Even though 
the evolution captured by the experiment between $\xi=0$ and
$\xi=0.0156$ is much less pronounced than the one evidenced 
by numerical simulations of Fig. 1, it should be emphasized that 
it is nevertheless significant and in reasonably good agrement 
with the evolution predicted by Eq. (\ref{Riem_approx}).
Indeed, the dashed black
lines in Fig. 3(c) represent the result of the numerical
integration  of Eq. (\ref{Riem_approx}) between $\xi=0$ and
$\xi=0.0156$ while starting from  initial conditions recorded in the
experiment. The  obtained
agreement between experiments and numerical simulations is acceptable
without being perfect because of limited signal to noise ratio in the
measurement  of $\rho$ and $u$.

\begin{figure}[h]
\includegraphics[width=0.45\textwidth]{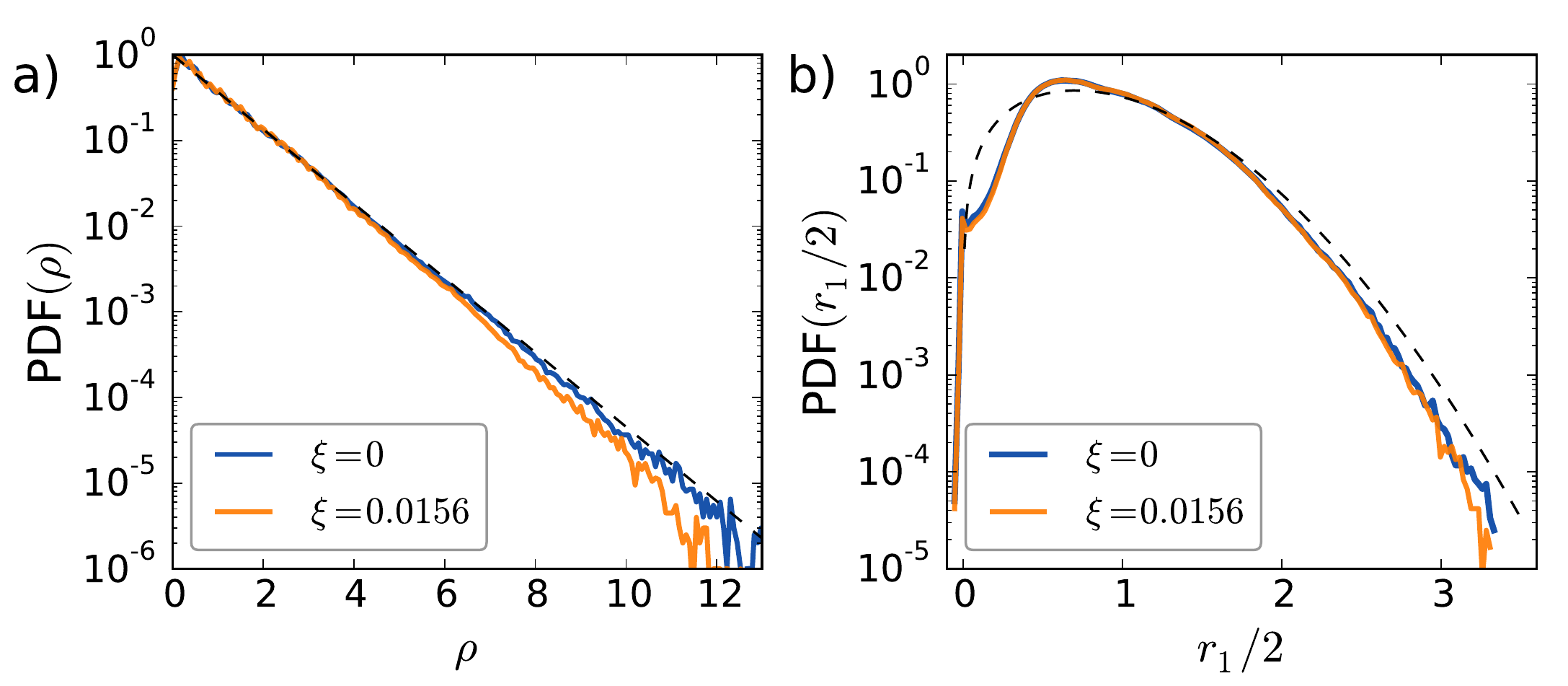}
\caption{Experimental PDFs of (a) $\rho$, 
  (b) $r_1/2$. Blue (resp. yellow)
  lines represent PDFs at the input (resp. output) end of the
  fiber. In (a) the black dashed line represents the exponential
  distribution and in (b), it represents
  the Rayleigh distribution.}
\end{figure}

Statistical features typifying dispersive-hydrodynamic IT 
have been investigated from 
long time series lasting $500 \mu$s and including
$80.10^{6}$ points. As shown in Fig. 4(a), the evolution depicted by the PDF
of $\rho$ is qualitatively similar to the one evidenced in Fig. 1(c)
and also in ref. \cite{Randoux:14}. On the other hand, as implied 
by the approximate 
decoupled RW system (\ref{Riem_approx}) the PDF of the Riemann
invariant $r_{1}/2$ practically does not change with $\xi$: it is
found to nearly retain the shape of the (initial) Rayleigh
distribution (note that initially $R_{1,2} \approx \pm 2 \sqrt{\rho}$), see Fig. 4(c). Note that 
$\sim 2\%$ of the measured points were excluded from the statistical 
analysis giving the PDFs of $u$ and $r_{1}/2$. For those points, the value
of $\rho$ is indeed too small for the proper determination of $u$. 

In conclusion, we have examined  the development  of IT from the
perspective of dispersive hydrodynamics. Within this framework the
initial stage of the IT development is described by a system of two
interacting random Riemann waves. Our analysis provides  an elementary
theoretical explanation of the fundamental IT phenomenon of the
appearance of low tails in the PDF of the wave's intensity. We  have
also shown from an optical fiber experiment that  
Riemann invariants represent  observable quantities
that provide new insight into the  description and the understanding
of IT. We hope that the dispersive-hydrodynamic approach used in
our work will pave the way to further theoretical  and experimental
studies in this field.

This work has been partially supported 
by the Agence Nationale de la
Recherche  through the LABEX CEMPI project (ANR-11-LABX-0007),  
the Ministry of Higher Education and Research,
Hauts de France council and European Regional Development 
Fund (ERDF) through the
the Nord-Pas de Calais Regional Research Council and the European
Regional Development Fund (ERDF) through the Contrat de Projets
Etat-R\'egion (CPER Photonics for Society P4S).






\end{document}